\documentclass[aps,pra,superscriptaddress,preprintnumbers,twocolumn]{revtex4}
\usepackage{natbib}
\usepackage{epsfig}
\usepackage{bm,epstopdf}
\usepackage{graphicx,times}
\usepackage[usenames]{color}
\usepackage[dvipsnames]{xcolor}
\usepackage{amssymb,amsmath,bbm,amsfonts,amsthm,subfigure}

\newcommand{\ket}[1]{|{#1}\rangle}
\newcommand{\bra}[1]{\langle{#1}|}

\newcommand{\exval}[1]{\langle{#1}\rangle}

\newcommand{\Tr}{{\rm Tr}}
\newcommand{\rhoin}{\rho^{\rm in}}
\newcommand{\rhoout}{\rho^{\rm out}}

\newcommand{\Id}{\mathbb{I}}

\newcommand{\ops}[2]{{#1}^{({#2})}}
\newcommand{\proj}[2]{\ket{#1}\!_{_{#2}}\!\bra{#1}}

\graphicspath{{figure/}}

\begin{document}
\title{A dynamical model for Positive-Operator Valued Measures}

\author{A. De Pasquale}
\affiliation{Dipartimento di Fisica, Universit\`a di Firenze,
      Via G. Sansone 1, I-50019 Sesto Fiorentino (FI), Italy}
\affiliation{INFN Sezione di Firenze, via G.Sansone 1,
      I-50019 Sesto Fiorentino (FI), Italy}

\author{C. Foti}
\affiliation{Dipartimento di Fisica, Universit\`a di Firenze,
      Via G. Sansone 1, I-50019 Sesto Fiorentino (FI), Italy}
\affiliation{INFN Sezione di Firenze, via G.Sansone 1,
      I-50019 Sesto Fiorentino (FI), Italy}

\author{A. Cuccoli}
\affiliation{Dipartimento di Fisica, Universit\`a di Firenze,
       Via G. Sansone 1, I-50019 Sesto Fiorentino (FI), Italy}
\affiliation{INFN Sezione di Firenze, via G.Sansone 1,
     I-50019 Sesto Fiorentino (FI), Italy}

\author{V. Giovannetti}
\affiliation{NEST, Scuola Normale Superiore and Istituto Nanoscienze-CNR, I-56127 Pisa, Italy}

\author{P. Verrucchi}
\affiliation{Istituto dei Sistemi Complessi,
       Consiglio Nazionale delle Ricerche,
       via Madonna del Piano 10,
       I-50019 Sesto Fiorentino (FI), Italy}
\affiliation{Dipartimento di Fisica, Universit\`a di Firenze,
       Via G. Sansone 1, I-50019 Sesto Fiorentino (FI), Italy}
\affiliation{INFN Sezione di Firenze, via G.Sansone 1,
       I-50019 Sesto Fiorentino (FI), Italy}

\begin{abstract}
 
We tackle the dynamical description of the quantum measurement process, by explicitly addressing the interaction between the system under investigation with the measurement apparatus, the latter ultimately considered as macroscopic quantum object. We consider arbitrary Positive Operator Valued Measures (POVMs), such that the orthogonality constraint on the measurement operators is relaxed. We show that, likewise the well-known von-Neumann scheme for projective measurements, it is possible to build up a dynamical model holding a unitary propagator characterized by a single time-independent Hamiltonian. This is achieved by modifying the standard model so as to compensate for the possible lack of orthogonality among the measurement operators of arbitrary POVMs.

\end{abstract}

\date{\today}

\maketitle 

\section{Introduction} \label{section1} 
 A distinctive trademark of quantum mechanics is represented by the quantum measurements 
and by the randomness of their outcomes. The postulates of the theory dictate how to compute the associated statistics for quantum observables through projective measures, 
while no mechanism is provided to predict how the actual finally observed result comes about. In this respect the measurement process still represents an open field of research~\citep{Holevo82, Davies1976,Kraus1983,BuschLM96, 
deMuynck02, Zurek03, Zurek81, HeinosaariZ12}. Actually, from the dawn of quantum theory, two main steps towards complementary directions have been performed. On the one hand a clear description of the measurement process entailing the definition of a time-independent interaction Hamiltonian between the system and an ultimately macroscopic apparatus has been provided by von Neumann~\citep{Neuman1927}, and fully characterized by Ozawa~\citep{Ozawa84} several years later. On the other hand, the statistical description of quantum measurements has been extended to non-necessarily orthonormal measurement operators by the introduction of  the so-called Positive-Operator Valued Measures (POVMs)~\citep{Holevo82,Davies1976,Kraus1983}. 

In this manuscript we unify these two approaches, introducing a dynamical description of arbitrary quantum measurements. We show that, in order to achieve a well-defined -- i.e., completely positive trace preserving 
(CPT)~\cite{Nielsen2004,Holevo82} -- dynamical map, the lack of orthogonality of arbitrary measurement operators needs to be compensated by properly
modifying the von Neumann-Ozawa (vN-O), time-independent Hamiltonian representation.
In our analysis we rely on the  Naimark extension theorem~\citep{Naimark40, 
Naimark43,Akhierzer1961, Paris12}, which allows one to describe an arbitrary POVM performed on the system of interest, in terms of a projective measurement performed on an external probing system  that was properly coupled with the latter. 
  This provides a proper generalization of the von Neumann model to arbitrary measurements. We recall that addressing the actual dynamics behind the formal 
description of a quantum measurement not only helps us understanding fundamental aspects of the process, but it also 
gives relevant indication about the actual design of quantum-measurement experiments (see, e.g., Refs.~\cite{BuffoniSVCC18,Binder19,DallaPozza}).

The paper is structured as follows: as a premise in Sec.~\ref{Sec2} we introduce the notation and 
review few basic notions regarding POVMs and the vN-O construction for projective measurements. 
Section~\ref{sec.constr} contains the original part of the work: Here we rigorously define the problem we wish to address and present a solution for it; the fundamental element of our
analysis is the explicit construction of a Naimark Hamiltonian discussed in Secs.~\ref{secFIRSTex}-\ref{secSECONDex}. 
Conclusion and final remarks are given in Sec.~\ref{Sec:conclusion}, while  
Technical considerations are presented in the Appendix. 

\section{Quantum Measurements} \label{Sec2}  The minimal description of a quantum measurement requires two 
elements: a set of $n_\Gamma$ distinguishable outcomes, $\{\mu_\gamma; \gamma=1,...n_\Gamma\}$, 
 and the corresponding probability distribution 
$\{p_\gamma\}$. Herewith, without loss of generality, we will exclusively consider countable sets of outputs 
and hence discrete distributions. 
This process involves at least two players interacting with each other: 
the system $S$, upon which the measurement is performed, and 
the apparatus $\Xi$, from which one actually obtains the outcomes.
Let ${\cal H}_{_{S}}$ be the Hilbert space of $S$. Formally, a quantum measure on a state $\rhoin_{_{S}}$
is defined by a bijection from 
$\{\mu_\gamma\}$ 
into the set of positive operators $\{ {F}^{(\gamma)}_{_{S}}\}$ on ${\cal H}_{_{S}}$, 
called {\it elements of the measure} or {\it effects}, such that 
$p_\gamma=\Tr[\rhoin_{_{S}}{F}^{(\gamma)}_{_{S}}]$, $\forall \gamma$. In order for 
$\sum_\gamma 
p_\gamma=1$ to hold, it must be $\sum_\gamma{F}^{(\gamma)}_{_{S}}=\Id_{_{S}}$.
As a process on $S$, a single measurement acts on an input state 
$\rhoin_{_{S}}$, upon which we want to gain some information, and produces 
one output $\mu_{\bar\gamma}$ with probability $p_{\bar\gamma}$, as defined 
above. 
After the interaction with the apparatus $\Xi$ and before the 
production of the outcomes, the system is described by the so-called  
{\it post-measurement} state $\rhoout_{_{S}}$ of $S$, defined as 
\begin{eqnarray}
\rhoout_{_{S}}&=&
\sum_\gamma M_{_{S}}^{(\gamma)}\rhoin_{_{S}} {M_{_{S}}^{(\gamma)}}^\dagger
=\sum_\gamma p_\gamma \ops{\rho}{\gamma}_{_{S}}~,
\label{e.rhoout}
\\
\rho_{_{S}}^{(\gamma)}&=&\frac{1}{p_\gamma}{M_{_{S}}^{(\gamma)}}\rhoin_{_{S}} 
{M_{_{S}}^{(\gamma)}}^\dagger,~p_\gamma=\Tr \Big[{M_{_{S}}^{(\gamma)}}\rhoin_{_{S}} {M_{_{S}}^{(\gamma)}}^\dagger\Big].
\label{e.rho-gamma-out}
\end{eqnarray} 
In this expression the operators $M_{_{S}}^{(\gamma)}$, dubbed as {\it measurement} or
{\it detection} operators, are
defined by ${F}^{(\gamma)}_{_{S}}=M_{_{S}}^{(\gamma)^\dagger}\ops{M}{\gamma}_{_{S}}$,
a decomposition always allowed, due to the positiveness of the 
measurement operators. We will  refer to the states 
$\ops{\rho}{\gamma}_{_{S}}$ as $\gamma$-{\it detected states}.
Notice that, in general,  $n_\Gamma$ is not constrained by the dimension of the Hilbert space $n_{_{S}}=\dim {\cal H}_{_{S}}$. This is because neither the elements $F_{_{S}}^{(\gamma)}$ of the POVM nor the measurement operators $M_{_{S}}^{(\gamma)}$ are required to satisfy any orthogonality constraint. This is actually the case for a more specific
type of quantum measurement  
defined by 
a Projective-Valued Measure (PVM), or projective 
measure. The latter is characterized by a set of operators $\Pi_{_{S}}^{(\gamma)}$ being 
orthonormal projectors on $S$, which implies that $n_\Gamma \leq n_{_{S}}$.
A PVM $\{ \Pi_{_{S}}^{(\gamma)}\}$ defines self-adjoint operators 
$O_{_{S}}=\sum_\gamma o_\gamma \Pi_{_{S}}^{(\gamma)}$, with $o_\gamma$ real 
$\forall\gamma$ and in one-to-one relation with $\mu_\gamma$ via an 
invertible {\it calibration} function 
$f(o_\gamma)=\mu_\gamma$ \cite{BuschLM96}.
In fact, the usual formulation of the quantum-measurement 
postulate refers to the above operators as ``observables'' and 
assigns the probability 
$p_\gamma=\Tr[\rhoin_{_{S}} \Pi_{_{S}}^{(\gamma)}]$ to the 
eigenvalue $o_\gamma$. As for the $\gamma$-detected states, their definition as
$\{\ops{\rho}{\gamma}_{_{S}}
=\Pi_{_{S}}^{(\gamma)}\rhoin_{_{S}}\Pi_{_{S}}^{(\gamma)}/p_\gamma\}$  is an 
integral part of the postulate for PVM, asserting that, after 
one single measurement with output $\mu_{\bar\gamma}$, the system is in 
the state $\rho_{_{S}}^{\bar\gamma}$ with absolute certainty . 
This gives the state $\rhoout_{_{S}}$
the consistent meaning of statistical mixture of the detected states
produced in a series of many identical repetitions of the measurement.
When rank$[\Pi_{_{S}}^{(\gamma)}]=1, \forall\gamma$, i.e. 
$\Pi_{_{S}}^{(\gamma)}=\proj{\gamma}{S}$, the PVM is called {\it ideal}, 
and $n_\Gamma={\rm dim}{\cal H}_{_{S}}$.

\subsection{The Naimark extension theorem} \label{naimark} 
\begin{figure}[h!] 
\includegraphics[width=0.25\textwidth]{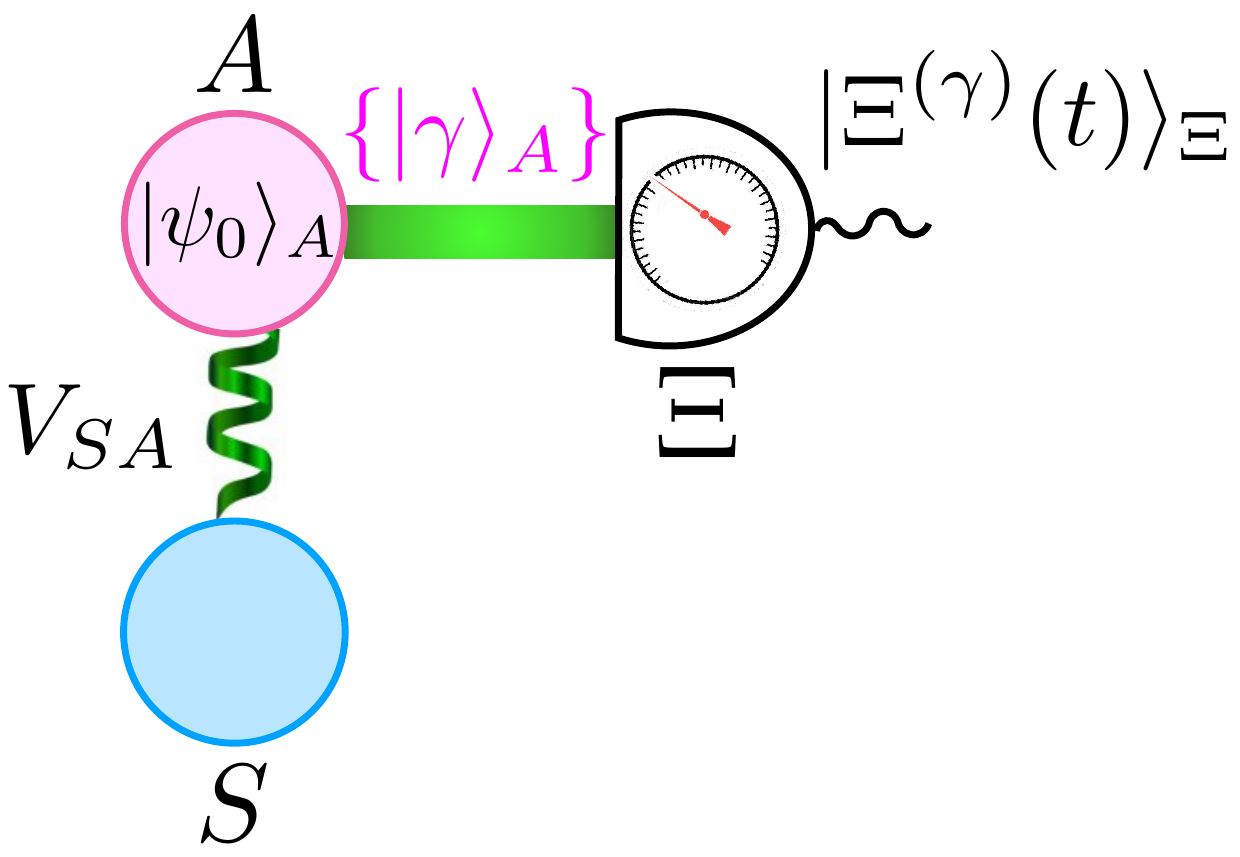} \caption{Scheme of 
Naimak representation for a POVM $\{F^{(\gamma)}_{_S}\}$ on $S$, 
$F^{(\gamma)}_{_S}=\mathrm{Tr}_A\left[\mathbb{I}_{_S}
\otimes\proj{0}{A}\;V_{_{S\!A}}^\dagger 
(\mathbb{I}_{_S}\otimes\proj{\gamma}{A})V_{_{S\!A}}\right]$.} 
\label{fig:Naimark} 
\end{figure}

The Naimark extension theorem~\citep{Naimark40, 
Naimark43,Akhierzer1961, Paris12} establishes a formal connection between POVM and PVM. 
Specifically it states 
that any given POVM $\{\ops{F}{\gamma}_{_{S}}\}$ for $S$ can be represented as a 
PVM $\{\ops{\Pi}{\gamma}_{_A}\}$ for an ancillary system $A$, that has unitarily
interacted with $S$ prior to be tested. Let $n_{_A}$ be the dimension of the Hilbert space ${\cal H}_{_A}$ associated to the ancilla. Formally, the Naimark theorem requires that 
$n_{_A}\ge n_\Gamma$, allowing the choice
$n_{_A}=n_\Gamma$ that entails an ideal PVM on $A$,
$\ops{\Pi}{\gamma}_{_A}=\proj{\gamma}{A}$. 
It then states that there exists: {\it i)} a state $\proj{\psi_0}{A}\in {\cal L}({\cal H}_{_{A}})$, 
{\it ii)} a unitary operator 
$V_{_{S\!A}}\in {\cal L}({\cal H}_{_{SA}})$ and {\it iii)} an ideal PVM $\{\proj{\gamma}{A}\}$ 
for $A$ (see Fig.~\ref{fig:Naimark}), such that 
\begin{eqnarray} 
\ops{F}{\gamma}_{_{S}} =\Tr_{_A}\left[(\mathbb{I}_{_{S}}\otimes\proj{\psi_0}{A})V^\dagger_{_{S\!A}}(\mathbb{I}_{_{S}}\otimes\proj{\gamma}{A})V_{_{S\!A}}\right] \;, 
\label{e.E_gamma-Naimark} 
\end{eqnarray} 
and
\begin{eqnarray} 
p_\gamma=\Tr\left[\left(\rhoin_{_{S}}\otimes\proj{\psi_0}{A}\right)
\left(V_{_{S\!A}}^\dagger(\mathbb{I}_{_{S}}\otimes\proj{\gamma}{A})V_{_{S\!A}} 
\right)\right]\,,
\label{e.p_gamma-Naimark}
\end{eqnarray}
which  allows us to consistently write
\begin{eqnarray} 
M_{_{S}}^{(\gamma)}=\!_{_{_A}}\!\bra{\gamma}V_{_{S\!A}}\ket{\psi_0}\!_{_{_A}}\label{DEFV}\;.\end{eqnarray} 

An explicit example of the above construction is presented in Sec.~\ref{secFIRSTex}: 
it should be stressed however that this is by no means the only possibility, as different 
choices for the Naimark operator  $V_{_{S\!A}}$ are typically available for each given POVM. 
It should also be noticed that, conversely, a unitary 
transformation of the state $\rhoin_{_{S}}\otimes\proj{\psi_0}{A}$ into 
$V_{_{S\!A}}(\rhoin_{_{S}}\otimes\proj{\psi_0}{A})V_{_{S\!A}}^\dagger$, followed by an 
ideal PVM $\{\proj{\gamma}{A}\}$ on $A$, defines a proper 
POVM on $S$. In this respect, the entrance of the 
ancilla is extremely valuable, as it provides the theoretical 
scheme with the versatility needed to describe diverse experimental 
situations, such as those where a physical mediator actually exists, 
and is ultimately responsible for the information transfer from $S$ to 
$\Xi$~\cite{PekolaJP15, Ronzani2018}.

\subsection{Dynamical models for PVM}  
\begin{figure}[h!]
\includegraphics[width=0.25\textwidth]{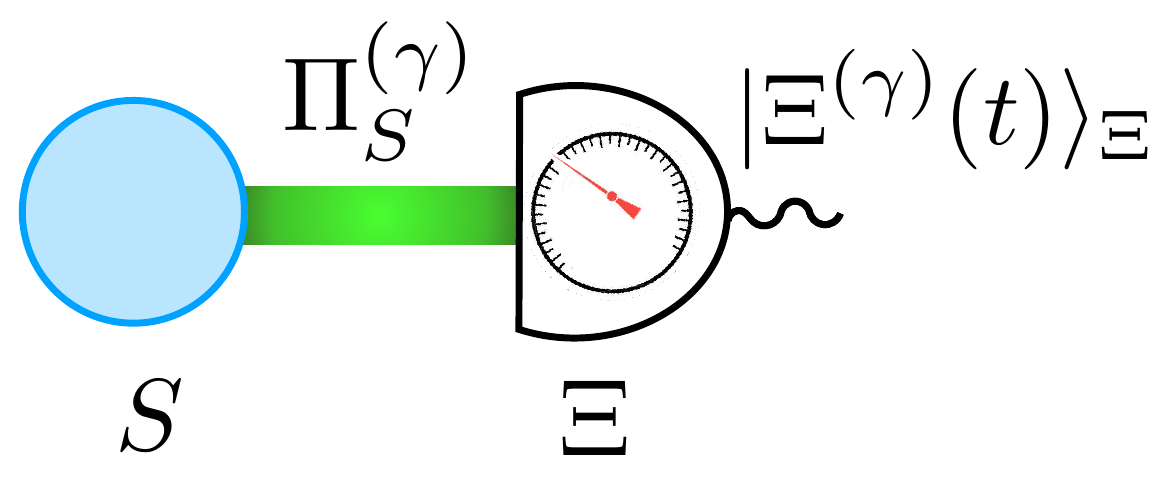}
\caption{Schematic representation of the von Neumann-Ozawa dynamical 
scheme for projective measure: an observable $O_{_S}=\sum_{\gamma} 
o_\gamma \Pi^{(\gamma)}_{_S} $ is measured on 
$S$ by letting it interact with the measurement apparatus $\Xi$, so as to 
encode the information on the states of the apparatus 
$\ket{\ops{\Xi}{\gamma}(t)}_{_\Xi}$.}
\label{fig:vonNeumann}
\end{figure}

Dynamical models for quantum measurements are meant to describe how a
measurement process takes place in time, in terms of a (time-independent) Hamiltonian coupling between
the system $S$ and an external environment $\Xi$ playing the role of the apparatus which, at the end of the process will store the measurement outcomes. More specifically, 
in its simplest, yet completely general version, the von Neumann-Ozawa (vN-O) dynamical model for 
PVM~\cite{Neuman1927,Wigner52,ArakiY60,Yanase61,ShimonyS79,Ozawa84}, assumes 
that the interaction between $S$ and $\Xi$ reads 
\begin{eqnarray}
\label{HAMVNO} 
H_{_{S\Xi}}:=O_{_{S}}\otimes O_{_{\Xi}}\;,\end{eqnarray} with $O_{_{S}} = \sum_{\gamma} o_{\gamma} \Pi_{_{S}}^{(\gamma)}$ an observable on $S$ and $O_{_{\Xi}}$ a 
self-adjoint  operator on $\Xi$ which  is canonically 
conjugated to what is typically referred to as ``the pointer" 
observable~\cite{Zurek81}.
Hence, the associated unitary evolution  writes
\begin{equation}\label{eq:vonNeumann_Ozawa}
U_{_{S\Xi}}(t) := e^{-i tO_{_{S}} \otimes O_{_{\Xi}} }= \sum_\gamma
 \Pi_{_{S}}^{(\gamma)}\otimes U^\gamma_{_{\Xi}}(t)~,
\end{equation}
where $U^\gamma_{_{\Xi}}(t)= e^{-i t o_\gamma O_{_{\Xi}} }$, in units $\hbar=1$.
 The model also assumes that
$\Xi$ is initially prepared in a pure state 
$\ket{D}$ that is not an eigenstate of $O_{\Xi}$.
If the system $S$ is initialized at $t=0$ in the state 
 $\rhoin_{_{S}}$, the unitary \eqref{eq:vonNeumann_Ozawa} 
makes the system  $S+\Xi$ evolve into the joint density matrix 
\begin{equation}\label{e.maprho_{_{S}}(t)1}
\rho_{_{S\Xi}}(t):=\sum_{\gamma\gamma'}\ops{\Pi}{\gamma}_{_{S}}\rhoin_{_{S}}\ops{\Pi} {\gamma'}_{_{S}} \otimes
|\Xi^\gamma(t) \rangle_{\Xi}\langle \Xi^{\gamma'}(t)|~,
\end{equation}
which upon partial trace with respect to $\Xi$, corresponds to the following local mapping 
\begin{equation}\label{e.maprho_{_{S}}(t)}
\rho_{_{S}}(t) =
\sum_{\gamma\gamma'}\ops{\Pi}{\gamma}_{_{S}}\rhoin_{_{S}}\ops{\Pi} {\gamma'}_{_{S}}
\,_{_{_{\Xi}}}\!\exval{\Xi^{\gamma'}(t)|\Xi^\gamma(t)}_{_{_{\Xi}}}~,
\end{equation}
 for $S$.
In the above expressions 
$\ket{\Xi^\gamma(t)}_{_{\Xi}}:=U^\gamma_{_{\Xi}}(t)\ket{D}_{_{\Xi}}$ are pure states of $\Xi$  which encode the
measurement outcomes $\gamma$ (see Fig.\ref{fig:vonNeumann} for a schematic representation of the process). 
The more distinguishable are such states, the larger is the information stored in $\Xi$  that allows one to 
distinguish between the different outcomes. 
In fact, the most favourable 
situation in terms of information transfer from $S$ to 
$\Xi$, corresponds to have the $\ket{\Xi^\gamma(t)}_{_{\Xi}}$s 
orthonormal. 
It is easily seen that when 
this condition holds, from Eq.~(\ref{e.maprho_{_{S}}(t)})  it follows that the matrix-representation of $\rho_{_{S}}(t)$ on the 
basis of the $O_{_{S}}$ eigenstates is block diagonal, and viceversa, i.e.
$\rho_{_{S}}(t)=\sum_{\gamma}\ops{\Pi}{\gamma}_{_{S}}\rhoin_{_{S}}\ops{\Pi} {\gamma}_{_{S}}$, 
 as required by \eqref{e.rhoout}  if $M^{(\gamma)}_{_{S}}=\Pi^{(\gamma)}_{_{S}}=F^{(\gamma)}_{_{S}}$.
This clarifies why decoherence plays such an important role in the 
quantum measurement process
~\citep{NamikiPN97,Schlosshauer07,BellomoDPGM10,LiuzzoScorpoCV15,Foti2018}. 
Therefore we say that the PVM $\{\ops{\Pi}{\gamma}_{_{S}}\}$ can be successfully realized on $S$ 
only if, in the limit of a macroscopic apparatus \citep{LiuzzoScorpoCV15,Foti2018}, it exists a time $t_{\rm d}$, typically referred to as {\em decoherence 
time}~\citep{Zurek81,Zurek03}, such that for $t > t_{\rm d}$ one has
\begin{eqnarray} 
_{_{_{\Xi}}}\!\exval{\Xi^{\gamma'}(t)|\Xi^\gamma(t)}_{_{_{\Xi}}} = \delta_{\gamma\gamma'}\label{MUTUALLY}\;,\end{eqnarray} 
or, at least, such that the above condition is approximately verified over some non trivial time interval preceding the
data acquisition event  (notice that although these scalar products are in principle periodic functions of 
time, in the 
limit of a macroscopic measuring device $\Xi$ one can safely take the time 
during which they stay approximately null much longer than the time necessary to 
perform the measurement). 

\section{Dynamical model for arbitrary POVM}  \label{sec.constr} 
In this section we discuss how to generalize the vN-O construction for PVM to the case of
 arbitrary POVMs, removing the constraint on the orthonormality of the measurement operators.
More precisely we show how to modify Eqs.~(\ref{HAMVNO}) and (\ref{eq:vonNeumann_Ozawa}) 
in such a way that for times $t$ larger than a certain 
characteristic threshold time $t_{\rm d}$, the interaction between $S$ and  $\Xi$  will yield 
a joint density matrix of the form similar to Eq.~(\ref{e.maprho_{_{S}}(t)1}), i.e.
\begin{equation}\label{e.maprho_{_{S}}(t)1POVM}
\rho_{_{S\Xi}}(t)=\sum_{\gamma\gamma'}\ops{M}{\gamma}_{_{S}}\rhoin_{_{S}}{\ops{M} {\gamma'}_{_{S}}}^\dag \otimes
|\Xi^\gamma(t) \rangle_{\Xi}\langle \Xi^{\gamma'}(t)|~,
\end{equation}
where $\{\ops{M}{\gamma}_{_{S}}; \gamma=1,\cdots,n_\Gamma\}$ are elements of the POVM and where
the vectors $\{ |\Xi^\gamma(t) \rangle_{\Xi}; \gamma=1,\cdots,n_\Gamma\}$ form a mutually orthonormal set as in Eq.~(\ref{MUTUALLY}).

Let us start by observing that at variance with the PVM scenario discussed in the previous section, 
we cannot expect Eq.~(\ref{e.maprho_{_{S}}(t)1POVM}) to apply at 
 those times  $t<t_{\rm d}$ for which
Eq.~(\ref{MUTUALLY}) does not hold.  
Indeed 
 due to the lack of orthogonality of the operators 
$\ops{M}{\gamma}_{_{S}}$ in this regime  
the resulting transformation  would not be  CPT in general, hence non physically implementable -- see Appendix
\ref{appax}.
This of course  does not imply that dynamical models cannot be found that describe a generic POVM: 
simply we need to  replace the  vN-O Hamiltonian coupling (\ref{HAMVNO}) with something else. 
The key ingredient for this construction is clearly provided by the  
Naimark extension theorem~\citep{Naimark40, 
Naimark43,Akhierzer1961, Paris12} we reviewed in Sec.~\ref{naimark}, which could be pictorially summarized as in Fig.~\ref{fig:Naimark}.
A tentative idea would be to work in a $S+A+\Xi$ scenario with a conventional vN-O couplings linking the 
apparatus $\Xi$ to $A$ or to $S+A$ ($A$ being the Naimark ancillary system). However, this approach, which we briefly review in Appendix~\ref{sec:naim2}, does not conclusively work,
as, while being able to reproduce the correct outcome probability distribution, it cannot yield a solution capable to  approach Eq.~(\ref{e.maprho_{_{S}}(t)1POVM}) at some future time. 
On the contrary, a simpler and  more effective way to construct a dynamical model for 
an arbitrary POVM is found by identifying the system environment $\Xi$ directly with  $A$. 
Under this assumption  we then look for  
a proper Hamiltonian   coupling $H_{_{SA}}$ generating an unitary evolution $U_{_{SA}}(t) := e^{-it H_{_{SA}}}$
which for all times $t$ larger than a certain critical time $t_{\rm d}$ fulfills, at least approximately, the constraint 
\begin{eqnarray}  \label{DD} 
U_{_{SA}}(t) = V_{_{SA}} \;,
\end{eqnarray}  
$V_{_{SA}}$ being the unitary entering Eq.~(\ref{DEFV}). 
Clearly due to the Stone theorem~\cite{STONE,Messiah} such a Naimark Hamiltonian can always be 
identified.
However, our goal is to produce an explicit construction for such a term, as we show in the 
following.

\begin{figure}[h!]
\includegraphics[width=0.3\textwidth]{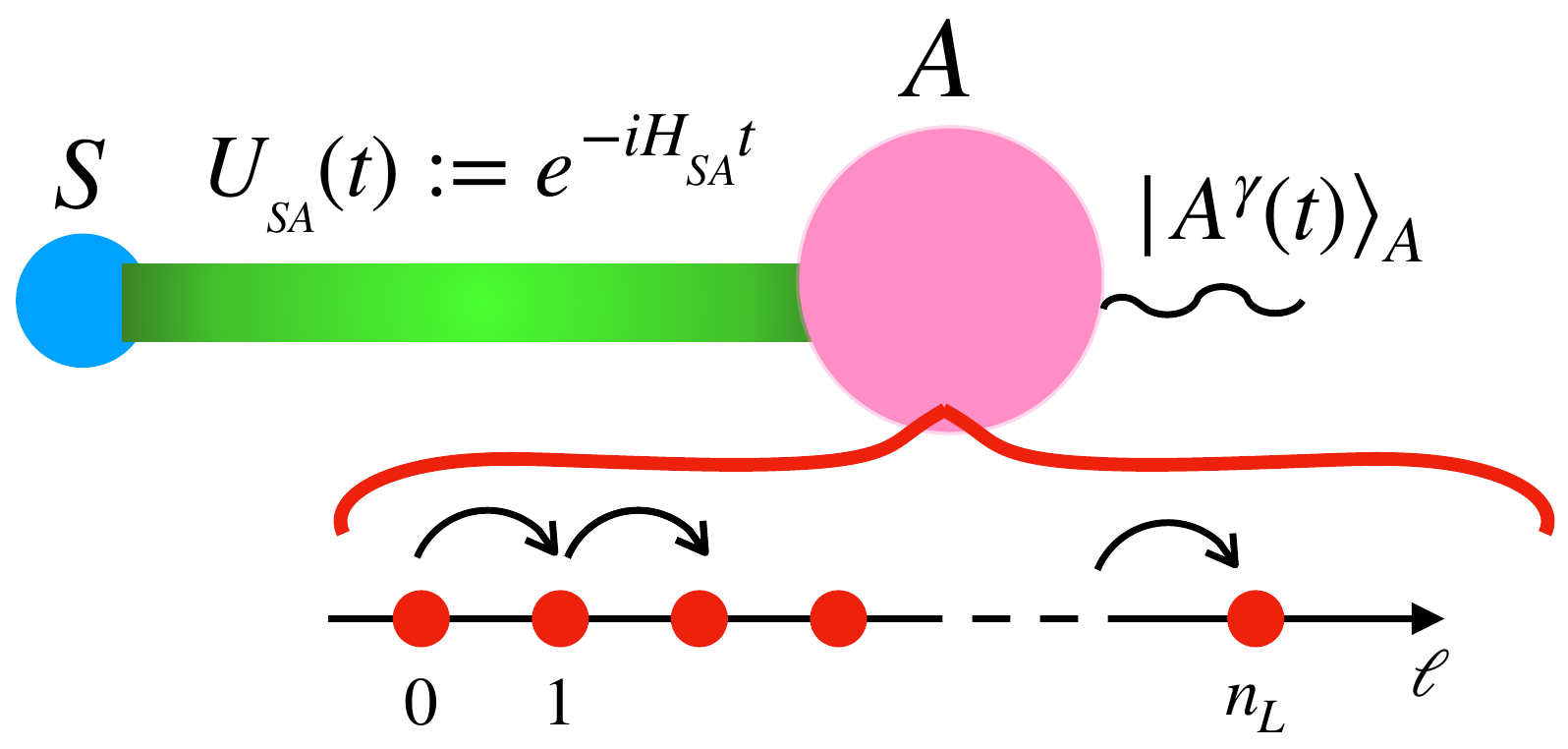}
\caption{Schematic representation of our dynamical model for POVMs. The principal system $S$ interacts with an ultimately macroscopic anicilla $A$. The $S+A$ coupling is ruled by a time independent generator $H_{_{SA}}$. In fact, the unitary transformation  $U_{_{SA}}(t)$ induces the transition from an arbitrary initial state $\rho_{_S}^{\rm in} \otimes |\psi_0\rangle_{_A}\langle \psi_0|$  to the final state  \eqref{e.maprho_{_{S}}(t)POVM3}. The information about the possible outputs $\mu_\gamma$ is encoded in the orthonormal states of ancilla $|A^{(\gamma)}(t)\rangle_{_A}$.  }
\label{fig:chain}
 \end{figure}

In order to construct our candidate for $H_{_{SA}}$ we start with a first example that 
utilizes a small ancilla $A$, hence inducing a $S+A$ dynamics which is explicitly periodic: accordingly 
this model is capable to produce the same correlations as in Eq.~(\ref{e.maprho_{_{S}}(t)1POVM}) 
only for specific values of $t$, with cyclic recurrence that prohibits the possibility of
maintaining such configuration indefinitely or at least for some non-zero  time intervals. 
 The second model, which is actually the central result of this manuscript, corrects this drawback adopting a much larger ancilla. A pictorial representation of the model is presented in Fig.\ref{fig:chain}, while the complete analytical derivation is presented in Sec.~\ref{secSECONDex}.
 In the specific, we introduce a degeneracy parameter $\ell=1, \ldots, n_L$ for the ancilla Hilbert space ${\cal H}_{_A}$ and define a coupling between $S$ and $A$ formally equivalent to first neighboring hopping terms, characterizing models for perfect state transfer~\cite{ChristandlPRL,ChristandlPRA}. Therefore, by increasing  $n_L$ it is possible to extend the condition Eq.~(\ref{e.maprho_{_{S}}(t)1POVM}) over arbitrarily large (ideally infinitely long) time intervals, as shown in Fig.~\ref{probability}. Actually, our model allows us to traslate~(\ref{e.maprho_{_{S}}(t)1POVM}) into
  \begin{equation}\label{e.maprho_{_{S}}(t)POVM3}
\rho_{_{SA}}(t)=\sum_{\gamma\gamma'}\ops{M}{\gamma}_{_{S}}\rhoin_{_{S}}{\ops{M} {\gamma'}_{_{S}}}^\dag \otimes
|A^{(\gamma)}(t) \rangle_{A}\langle A^{(\gamma')}(t)|~,
\end{equation}
where we have explicitly identified the state $|\Xi^{(\gamma)}(t)\rangle_{_\Xi}$ with the state of the enlarged ancilla $|A^{(\gamma)}(t)\rangle_{_A}$. A crucial difference between  $\{|\Xi^{(\gamma)}(t)\rangle_{_\Xi}\}$ and $\{|A^{(\gamma)}(t)\rangle_{_A}\}$ is that the latter are orthogonal to each other for all times $t$. This compensates for the possible lack of orthogonality of the measurement operators $M_{_S}^{(\gamma)}$, guaranteeing a posteriori the complete positivity of the unitary transformation $U_{_{SA}}(t)$.  

\subsection{First implementation: periodic dynamics} \label{secFIRSTex} 
Our first step to tackle the problem is to explicitly  write down a suitable candidate for the Naimark unitary $V_{_{SA}}$.
We observe that Eq.~(\ref{DEFV}) can be satisfied, e.g. by
requiring that for all $\ket{\psi}_{_{S}}$ of $S$
  the following
condition holds, 
\begin{eqnarray} 
V_{_{SA}} |\psi\rangle_{_{S}} \otimes |\psi_0\rangle_{_A} = e^{i\alpha} \sum_{\gamma =1}^{n_\Gamma} 
M_{_{S}}^{(\gamma)} \ket{\psi}_{_{S}} \otimes \ket{{\gamma}}\!_{_{_A}}\label{DEFV1}\;,\end{eqnarray} 
with $\{\ket{{\gamma}}_{_A}; \gamma= 1,2,\cdots, n_\Gamma\}$ being the orthornormal set of vectors of
 $A$ entering Eq.~(\ref{DEFV}), the phase $\alpha$ being absolutely irrelevant but being inserted for future reference (notice that the above requirement is fully consistent with the dimension $n_{_A}$ of $A$ being larger
than the total number of measurements outcomes $n_\Gamma$).  
This transformation does not completely characterize $V_{_{SA}}$ on the full Hilbert space of $S+A$, but does it
only on a proper subspace of the latter -- specifically the subspace associated with vectors having $A$ into 
the input state $\ket{\psi_0}$. By construction, at least on these vectors, it preserves the scalar product:
hence it can be generalized to a global unitary acting on the full space of the system and of the ancilla.
What we are going to do next is to explicitly construct such extension using a simplifying trick. 
Specifically, we assume the input vector $\ket{\psi_0}_{_A}$ of $A$ to be orthogonal to all the elements of the orthonormal set  $\{ \ket{{\gamma}}_{_A}; \gamma=1,\cdots, n_\Gamma\}$, i.e.
\begin{eqnarray} 
{_{_A}\langle} \psi_0 | {\gamma}\rangle_{_A} = 0\;,  \quad \forall \gamma=1,2,\cdots,n_\Gamma \;.\end{eqnarray}
This, of course, automatically implies that the dimension of $A$ we are considering has to be at least larger than or equal to  
$n_\Gamma +1$, i.e. slightly larger than the minimum value required by the Naimark theorem (i.e. $n_\Gamma$). 
Such small overhead turns out to be extremely useful as we now can decompose the matrix $V_{_{SA}}$ of 
(\ref{DD}) into a collection of $2\times2$ independent blocks. Indeed, let us introduce an orthonormal basis
$\{ \ket{{j}}_{_{S}}; {j=1,\cdots, n_{_{S}}}\}$  for ${\cal H}_{_S}$. Expanding $\ket{\psi}_{_{S}}$  in such a basis we can then 
observe that the identity  (\ref{DEFV1}) gets replaced by
\begin{eqnarray} \label{DD1} 
V_{_{SA}}  |\xi_{j}^{(0)}  \rangle_{_{SA}} &=& e^{i\alpha} |\xi_{j}^{(1)} \rangle_{_{SA}} \;, \end{eqnarray} 
where for all $j= 1,\cdots, n_{_{S}}$ we defined the pure states 
\begin{eqnarray} 
|\xi_{j}^{(0)}  \rangle_{_{SA}} &:=& \ket{{j}}_{_{S}}\otimes\ket{\psi_0}_{_A}\;, \label{DEFXIJ0} \\ 
|\xi_{j}^{(1)} \rangle_{_{SA}} &:=& \sum_{\gamma =1}^{n_\Gamma} 
M_{_{S}}^{(\gamma)} \ket{{j}}_{_{S}}\otimes\ket{{\gamma}}_{_A}\label{DEFV2}\;.\end{eqnarray} 
that 
 by construction are all mutually orthonormal, i.e. 
\begin{eqnarray}   \label{ORTOORTO} 
 {_{_{SA}}\langle}  \xi_{j}^{(\ell)}   |\xi_{j'}^{(\ell')}  \rangle_{_{SA}} = \delta_{j,j'}  \delta_{\ell,\ell'} \;, 
 \end{eqnarray} 
 with $\ell, \ell' = 0,1$.
They can be grouped in a collection of $n_{_{S}}$ 
mutually orthogonal,  2-dimensional subspaces 
\begin{eqnarray} \label{DEFHSA} {\cal H}_{_{SA}}^{(j)}:=\mbox{Span}\{ 
 |\xi_{j}^{(0)}  \rangle_{_{SA}},|\xi_{j}^{(1)}  \rangle_{_{SA}}\}\;,\end{eqnarray}  
 labelled by $j$ and spanned by the couple  $|\xi_{j}^{(0)}  \rangle_{_{SA}}$ and $|\xi_{j}^{(1)} \rangle_{_{SA}}$.
According to  (\ref{DD1})   the unitary $V_{_{SA}}$ operates separately on each one of the ${\cal H}_{_{SA}}^{(j)}$ where, up to the global phase $e^{i\alpha}$, it acts     as 
 the following effective Pauli transformations 
\begin{eqnarray} 
\sigma_{_{SA}}^{(j)} = [\sigma_{_{SA}}^{(j)}]^\dag  :=  |\xi_{j}^{(0)}  \rangle_{_{SA}}\langle \xi_{j}^{(1)}|   + 
|\xi_{j}^{(1)} \rangle_{_{SA}}\langle \xi_{j}^{(0)} | \;,
\end{eqnarray} 
leading to the identification 
\begin{eqnarray} 
V_{_{SA}} \label{DECVSIGMA} 
&=&  e^{i\alpha} \oplus_{j}   \sigma_{_{SA}}^{(j)}   \;,
\end{eqnarray} 
the direct sum being performed over all $j=1,\cdots, n_{_{S}}$.
Our first choice for the Naimark Hamiltonian is hence provided by the self-adjoint operator
\begin{eqnarray} \label{HAMNAI} 
H_{_{SA}} : = {\omega} \sum_{j=1}^{n_{_{S}}} \sigma_{_{SA}}^{(j)}\;,
\end{eqnarray}  
with ${\omega}>0$ an arbitrary positive constant, which, using~(\ref{DEFXIJ0}) and (\ref{DEFV2}) can be equivalently expressed as
\begin{eqnarray}
H_{_{SA}} = {\omega} \; \sum_{\gamma=1}^{n_\Gamma} \left( M_{_{S}}^{(\gamma)} \otimes |\gamma\rangle_{_A}\langle \psi_0| + h.c.\right).
\end{eqnarray} 
 Its associated unitary evolution is periodic of period
$2\pi/{\omega}$ and reads as
\begin{eqnarray} 
U_{_{SA}}(t) &: =&  e^{-i H_{_{SA}}  t } 
= \oplus_{j}  e^{-i {\omega} t  \sigma_{_{SA}}^{(j)} } \nonumber\\ 
&=& \oplus_{j}  \Big[ \openone_{_{SA}}^{(j)} 
\cos({\omega} t)- i  \sigma_{_{SA}}^{(j)} \sin({\omega} t)\Big]  \;, \label{gg1} 
\end{eqnarray} 
where we used the property
\begin{eqnarray} 
\sigma_{_{SA}}^{(j)}  \sigma_{_{SA}}^{(j')}  
= \delta_{j,j'} \openone_{_{SA}}^{(j)}\;,
\end{eqnarray} 
with 
\begin{eqnarray} 
\openone_{_{SA}}^{(j)} :=  |\xi_{j}^{(0)}  \rangle_{_{SA}}\langle \xi_{j}^{(0)} |   + 
|\xi_{j}^{(1)} \rangle_{_{SA}}\langle \xi_{j}^{(1)}| \;,
\end{eqnarray} 
being the projection operator on ${\cal H}_{_{SA}}^{(j)}$. From Eq.~(\ref{gg1}) it then follows 
\begin{eqnarray}  \label{fffd0} 
 &&U_{_{SA}}(t)|\psi\rangle_{_{S}}\otimes |\psi_0\rangle_{_A}=  
  \cos(\omega t)  |\psi\rangle_{_{S}}\otimes |\psi_0\rangle_{_A} \\ \nonumber 
 &&\qquad \qquad -i \sin(\omega t)  \sum_{\gamma=1}^{n_\Gamma} 
 M_{_{S}}^{(\gamma)} |\psi\rangle_{_{S}} \otimes  | \gamma \rangle_{_A} \;,
 \end{eqnarray} 
which
 yields Eq.~(\ref{DEFV1}) for $t=t_{\rm d}=\pi/(2{\omega})$,  upon identifying the phase term $\alpha$ with  $-\pi/2$.

\subsection{Second implementation: non periodic dynamics}  \label{secSECONDex} 
The main drawback of the previous example is that it exhibits a definite period 
$2\pi /{\omega}$, so that Eq.~(\ref{fffd0}) reproduces Eq.~(\ref{DEFV1}) only at 
the precise instants $t_n=(2n+1)t_{\rm d}$, where $n$ is an integer 
number. Hence it does not exactly fits into our  requirement to enforce Eq.~(\ref{e.maprho_{_{S}}(t)1POVM})
for extended time interval after a given premeasurement time $t_{\rm d}$. 
Here we show however how one can easily modify the construction to explicitly fulfill this requirement too.
The idea is 
 to increase the dimension of  the subspaces {\color{blue}${\cal H}_{_{SA}}^{(j)}$} of Eq.~(\ref{DEFHSA}) and to equip the associated Hamiltonian with a reacher frequency spectrum.
For this purpose we replace the orthonormal set
$\{\ket{{\gamma}}_{_A}; \gamma= 1,\cdots, n_\Gamma\}$ entering the previous construction
  with a larger set of orthonormal vectors $\{\ket{{\gamma},\ell}_{_A}; \gamma= 1,\cdots, n_\Gamma\}$, the index $\ell$ being a degeneracy parameter which can take up to $n_L$ different values, i.e. 
  \begin{equation} 
 {_{_A} \langle} \gamma',\ell' \ket{{\gamma},\ell}_{_A} = \delta_{\gamma,\gamma'} \delta_{\ell,\ell'}\;,  \qquad 
 {_{_A} \langle}\psi_0 \ket{{\gamma},\ell}_{_A} = 0  \quad \forall \gamma,\ell \;,
  \end{equation} 
  which implicitly dictates that now $A$ must have a dimension $n_{_A}$ which is larger than or equal to  $n_\Gamma n_L +1$. 
With that in mind we then replace Eq.~(\ref{DEFHSA}) with the $n_L+1$ dimensional 
spaces
\begin{eqnarray} \label{DEFHSAnew} {\cal H}_{_{SA}}^{(j)}:=\mbox{Span}\{ 
 |\xi_{j}^{(0)}  \rangle_{_{SA}},|\xi^{(1)}_{j} \rangle_{_{SA}},
 \cdots , |\xi^{(n_L)}_{j} \rangle_{_{SA}}\}\;,\end{eqnarray}
with $|\xi_{j}^{(0)}  \rangle_{_{SA}}$ still defined as in Eq.~(\ref{DEFXIJ0}) and where, for $\ell=1, \cdots, n_L$, $|\xi^{(\ell)}_{j} \rangle_{_{SA}}$  are instead given by
\begin{eqnarray} 
|\xi^{(\ell)}_{j} \rangle_{_{SA}}  := \sum_{\gamma =1}^{n_\Gamma} 
M_{_{S}}^{(\gamma)} \ket{{j}}_{_{S}}\otimes\ket{{\gamma},\ell}_{_A}\label{DEFV2ell}\;,
\end{eqnarray} 
which still fulfill the orthogonality conditions~(\ref{ORTOORTO}).
Define hence the new self-adjoint operators
\begin{eqnarray} \label{HAMNAI1} 
H_{_{SA}}^{(j)} &:=&\sum_{\ell=0}^{n_L-1}  {\omega}_\ell \sigma_{_{SA}}^{(j,\ell)}\;,
\end{eqnarray} 
with ${\omega}_\ell>0$ being frequency terms that play the role of free parameters in the model and 
where,  for $\ell=0,\cdots, n_L-1$ the new Pauli operators $\sigma_{_{SA}}^{(j,\ell)}$ are given by 
\begin{equation} 
\sigma_{_{SA}}^{(j,\ell)} = [\sigma_{_{SA}}^{(j,\ell)}]^\dag 
:=  |\xi_{j}^{(\ell)} \rangle_{_{SA}}\langle \xi_{j}^{(\ell+1)}|   + 
|\xi_{j}^{(\ell+1)} \rangle_{_{SA}}\langle \xi_{j}^{(\ell)}| \;.
\end{equation}  
Notice that from the orthonormality conditions~(\ref{ORTOORTO}) it follows that, irrespectively from the values of $\ell$ and $\ell'$, the product of any two operators  $\sigma_{_{SA}}^{(j,\ell)}$ and $\sigma_{_{SA}}^{(j',\ell')}$ with $j\neq j'$ vanishes, i.e.
\begin{equation}\label{NEWHAM}  
\sigma_{_{SA}}^{(j,\ell)}  \sigma_{_{SA}}^{(j',\ell')}=0\,.
\end{equation}
Furthermore, the various $H^{(j)}_{_{SA}}$ terms have exactly the same matrix 
form with respect to the associated canonical basis of the associated spaces ${\cal H}_{_{SA}}^{(j)}$, i.e. 
\begin{eqnarray} \label{ISOSP} 
{_{_{SA}} \langle} \xi^{(\ell')}_{j} | H_{_{SA}}^{(j)}  |\xi^{(\ell)}_{j} \rangle_{_{SA}}=\omega_\ell (\delta_{\ell,\ell'+1} +
\delta_{\ell+1,\ell'} ) \;.
\end{eqnarray} 
Finally, we observe that $H_{_{SA}}^{(j)}$ formally corresponds to the 1-excitation sector of a spin-1/2 chain Hamiltonian, with open boundary
conditions, characterized by first neighbouring hopping terms, whose coupling terms are gauged by the frequencies $\omega_\ell$s.

We hence introduce as the new Hamiltonian of the $S+A$ system the operator
\begin{eqnarray} \label{HAMNAI1} 
H_{_{SA}} &: =&  \sum_{j=1}^{n_{_{S}}} H_{_{SA}}^{(j)}\;, 
\end{eqnarray}
which, making use of Eqs.~(\ref{DEFV2ell})  and (\ref{DEFXIJ0}), can be equivalently 
recasted in the following compact form 
\begin{eqnarray} \label{ALTHAM} 
H_{_{SA}}&=&  \sum_{\gamma=1}^{n_\Gamma}M_{_{S}}^{(\gamma)} \otimes \Theta_{_A}^{(\gamma)}
+\sum_{\gamma,\gamma'=1}^{n_\Gamma}  M_{_{S}}^{(\gamma)}
{M_{_{S}}^{(\gamma')}}^\dag 
 \otimes  \Theta_{_A}^{(\gamma,\gamma')}  \nonumber \\
 &&\qquad\qquad \qquad\qquad \qquad\qquad \qquad\qquad + h.c. 
\end{eqnarray} 
after defining the operators 
\begin{eqnarray}
\Theta_{_A}^{(\gamma)} &:=&  {\omega}_0   |\gamma,1 \rangle_{_A}\langle \psi_0| \;, 
\nonumber \\
 \Theta_{_A}^{(\gamma,\gamma')} &:=& \sum_{\ell=1}^{n_L-1} {\omega}_\ell  |\gamma,\ell \rangle_{_A}\langle \gamma', \ell+1| \;.
 \end{eqnarray} 
From Eqs.~(\ref{NEWHAM}) and (\ref{ISOSP}) 
we notice that, as in the case of Sec.~\ref{secFIRSTex}, $H_{_{SA}}$ is block diagonal,
with respect to the extended subspaces ${\cal H}_{_{SA}}^{(j)}$, with iso-spectral blocks, . 
Hence it acts independently on each one of such subspaces, inducing on each one of them the same local unitary rotation, i.e. 
\begin{eqnarray} 
U_{_{SA}}(t) &=&  e^{-i H_{_{SA}}  t } = \oplus_{j}  e^{-i t  H^{(j)}_{_{SA}} 
 } \;. 
\end{eqnarray} 
If we now consider the evolution it induces on an input state of the form $|\psi\rangle_{_{S}}\otimes |\psi_0\rangle_{_A}$, where  $|\psi\rangle_{_{S}}$ 
is a generic vector of $S$, expanding the input state 
 as a linear combination of the vectors $|\xi_{j}^{(0)} \rangle_{_{SA}}$, we can write 
 \begin{eqnarray}  \label{fffd} 
 U_{_{SA}}(t)|\psi\rangle_{_{S}}\otimes |\psi_0\rangle_{_A}=  \sum_{j=1}^{n_{_{S}}} \alpha_j |\xi_{j}^{(0)}(t) \rangle_{_{SA}} \;,
 \end{eqnarray} 
 $\alpha_j$ being the expansion coefficients of $|\psi\rangle_{_{S}}$ with respect to the basis $\{ |j\rangle_{_{S}};
 j=1,\cdots, n_{_{S}}\}$ and where the vector 
 \begin{eqnarray} \label{EQUA} 
 |\xi_{j}^{(0)}(t) \rangle_{_{SA}} :=  e^{-i t  H^{(j)}_{_{SA}}} |\xi_{j}^{(0)} \rangle_{_{SA}}\;,\end{eqnarray} 
 is the evolution of $|\xi_{j}^{(0)}(t) \rangle_{_{SA}}$ induced by the Hamiltonian component $H^{(j)}_{_{SA}}$ that
 is active on the subspace ${\cal H}_{S,A}^{(j)}$.
 By construction $|\xi_{j}^{(0)}(t) \rangle_{_{SA}}\in {\cal H}_{S,A}^{(j)}$ so that we can write it as 
 \begin{eqnarray} \label{ddd} 
 |\xi_{j}^{(0)}(t) \rangle_{_{SA}} = \sum_{\ell=0}^{n_L} \beta_\ell (t) |\xi_{j}^{(\ell)}  \rangle_{_{SA}}  \;.\end{eqnarray} 
In this expression the quantities $\beta_\ell(t)$ are (properly normalized) amplitude probabilities associated with the canonical orthonormal basis 
$|\xi_{j}^{(0)}  \rangle_{_{SA}},|\xi^{(1)}_{j} \rangle_{_{SA}},
\cdots , |\xi^{(n_L)}_{j} \rangle_{_{SA}}$, whose 
 explicit functional dependence on $t$ can be freely tailored by properly
 choosing  the
frequencies $\omega_1$, $\omega_2$, $\cdots$, $\omega_{n_L}$ of the model.
The relevant observation here is the fact that due to the iso-spectral property (\ref{ISOSP}),
such coefficients do not bear any functional dependence upon the index $j$.
Exploiting this fact and replacing Eq.~(\ref{ddd}) into (\ref{fffd}) we can hence write 
\begin{eqnarray}  
 &&U_{_{SA}}(t)|\psi\rangle_{_{S}}\otimes |\psi_0\rangle_{_A}=  
 \beta_0(t) |\psi\rangle_{_{S}}\otimes |\psi_0\rangle_{_A} \\ \nonumber 
 &&\qquad \qquad +\sqrt{1- |\beta_0(t)|^2} \sum_{\gamma=1}^{n_\Gamma} 
 M_{_{S}}^{(\gamma)} |\psi\rangle_{_{S}} \otimes  | A^{\gamma}(t) \rangle_{_A} \;,
 \end{eqnarray} 
where for $\gamma=1,\cdots, n_\Gamma$, 
\begin{eqnarray} 
| A^{\gamma}(t) \rangle_{_A}:= \frac{1}{\sqrt{1- |\beta_0(t)|^2} }  \sum_{\ell=1}^{n_L} \beta_\ell (t)
 |\gamma,\ell \rangle_{_A}\;,
\end{eqnarray} 
 form an orthonormal set of vectors of $A$ which are also orthogonal to $|\psi_0\rangle_{_A}$, i.e. they fulfill the conditions
 \begin{eqnarray} \label{orto11}
 {_{_A} \langle}  A^{\gamma'}(t) | A^{\gamma}(t) \rangle_{_A} = \delta_{\gamma,\gamma'}\;, \qquad
  {_{_A} \langle} \psi_{0} | A^{\gamma}(t) \rangle_{_A}  =0\;.
 \end{eqnarray} 

As a consequence of Eq.~(\ref{fffd}), it follows that the evolved density matrix $\rho_{_{SA}}(t):= U_{_{SA}}(t) (\rhoin_{_{S}} \otimes |\psi_0\rangle_{_A}\langle \psi_0|) U^\dag_{_{SA}}(t)$  of $S+A$ at time $t$ can be written as
\begin{eqnarray}\nonumber 
&& \rho_{_{SA}}(t) = |\beta_0(t)|^2 \rhoin_{_{S}} \otimes |\psi_0\rangle_{_A}\langle \psi_0|   + |\beta_0(t)| \Delta_{SA}(t) \nonumber \\
&+&  (1- |\beta_0(t)|^2) 
\sum_{\gamma\gamma'}\ops{M}{\gamma}_{_{S}}\rhoin_{_{S}}{\ops{M} {\gamma'}_{_{S}}}^\dag \otimes
| A^{\gamma}(t) \rangle_{_A}\langle A^{\gamma'}(t)| \;,  \nonumber \\ 
\label{e.maprho_{_{S}}(t)11example}
\end{eqnarray}  
where we have define the bounded operator on $S+A$
\begin{eqnarray} 
\Delta_{_{SA}}(t) 
&=&  e^{- i {\xi_0(t)}}  \sqrt{1-|\beta_0(t)|^2}
\sum_{\gamma}\ops{M}{\gamma}_{_{S}}\rhoin_{_{S}} \otimes | A^{\gamma}(t) \rangle_{_A}\langle\psi_{0}| \nonumber \\
&&\qquad \qquad +h.c. \;,
\end{eqnarray} 
$e^{i {\xi_0(t)}}$  being the phase of $\beta_0(t)$.
 
 \begin{figure}[t!]
\includegraphics[width=0.4\textwidth]{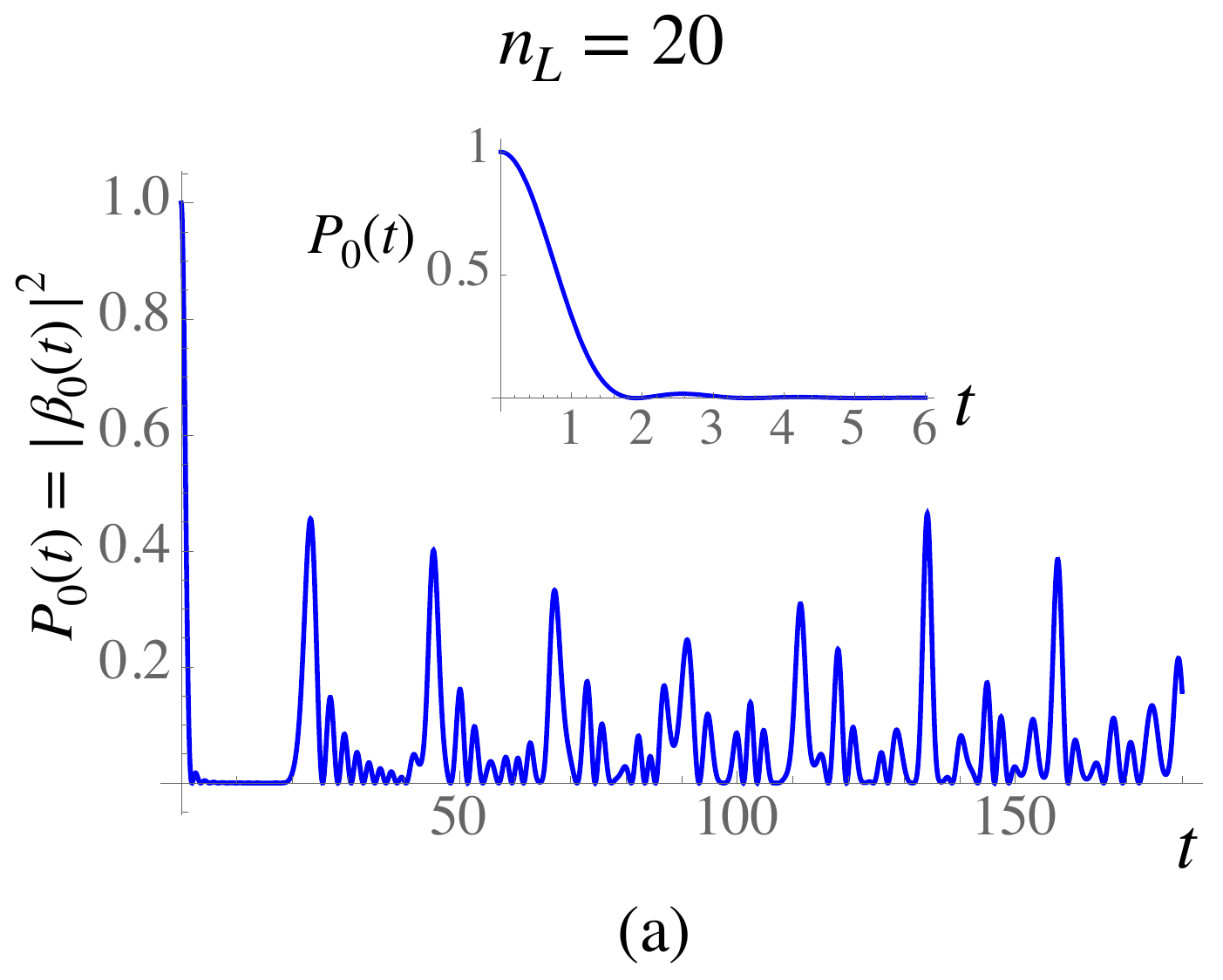}
\\ \vspace{0.6cm}
\includegraphics[width=0.4\textwidth]{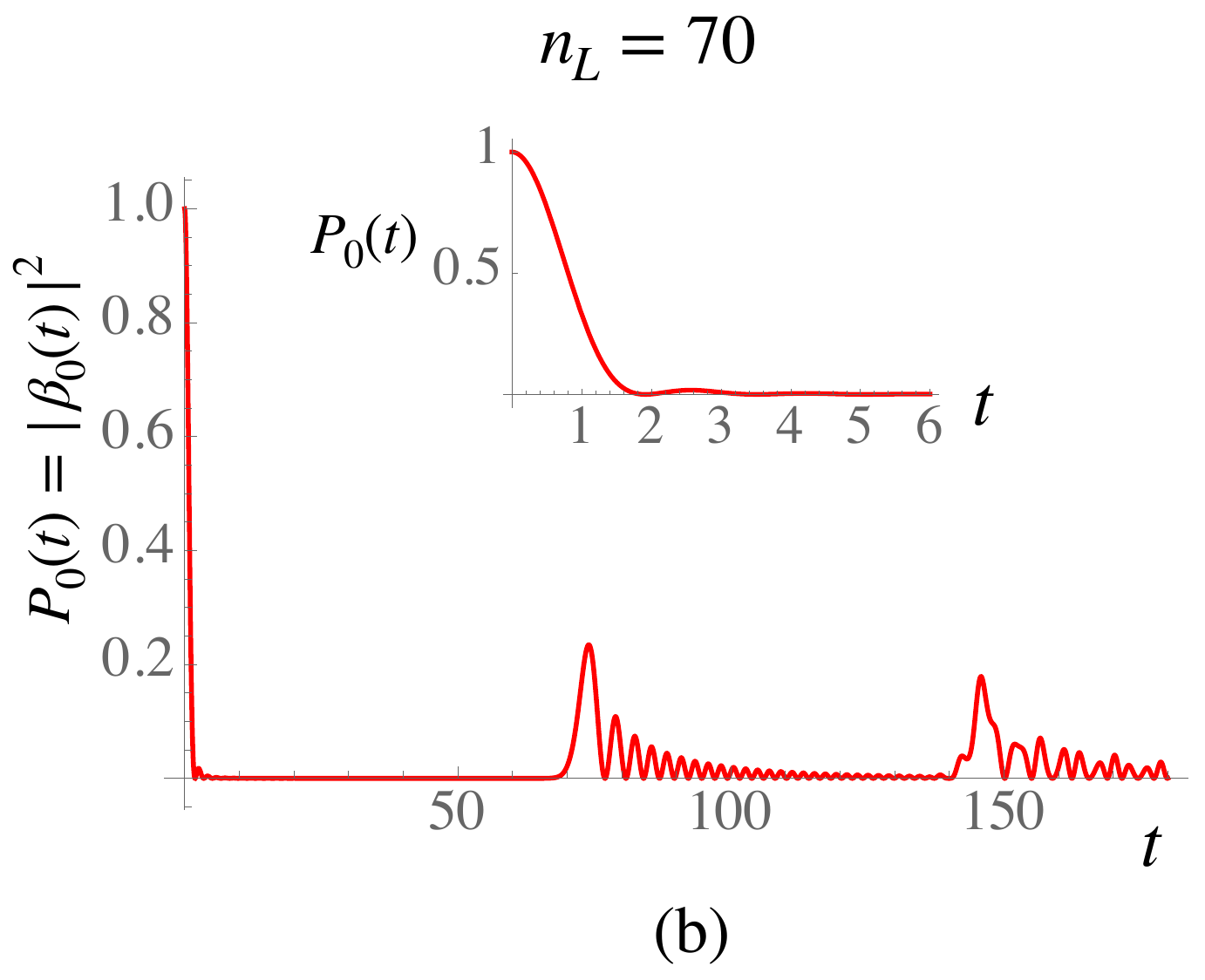}
\caption{Plot of the probability function $P_0(t):=|\beta_0(t)|^2$ entering Eq.~(\ref{e.maprho_{_{S}}(t)11example}) obtained by solving Eq.~(\ref{EQUA}) 
for  $n_L=20$ (panel (a)) and $n_L=70$ (panel (b)) for the case where the frequency parameters $\omega_\ell$ of Eq.~(\ref{ISOSP}) are taken to be uniform and equal to $\omega_0$. 
 In both cases $P_0(t)$ drops from $1$ to almost zero around $t\sim 2/\omega_0$. Then small revivals appear quite periodically after a time interval approximately given by $n_L$. Therefore, in the limit of $n_L\to \infty$, and after a given lapse of time (see the insets), the state of the $SA$ can be safely approximated by \eqref{e.maprho_{_{S}}(t)11example1}.}
\label{probability}
\end{figure}

The relevant quantity in Eq.~(\ref{e.maprho_{_{S}}(t)11example})  is the  probability amplitude function $\beta_0(t)$:  for $t=0$ it is equal to $1$, in agreement with the requirement that 
 $\rho_{_{SA}}(0)=\rhoin_{_{S}} \otimes |\psi_0\rangle_{_A}\langle \psi_0|$, but $\beta_0(t)\rightarrow 0$ in an extended time interval for large enough $n_L$, as shown in Fig.~\ref{probability}.  
 Accordingly, in such time interval the above expression reduces to 
  \begin{equation}\label{e.maprho_{_{S}}(t)11example1}
\rho_{_{SA}}(t)\simeq 
\sum_{\gamma\gamma'}\ops{M}{\gamma}_{_{S}}\rhoin_{_{S}}{\ops{M} {\gamma'}_{_{S}}}^\dag \otimes
| A^{\gamma}(t) \rangle_{_A}\langle A^{\gamma'}(t)|  \;,
\end{equation}  
which effectively achieves our target~(\ref{e.maprho_{_{S}}(t)1POVM}) by identifying 
$|A^{\gamma}(t) \rangle_{_A}$ with $|\Xi^\gamma(t) \rangle_{_{\Xi}}$.

\section{Conclusions}\label{Sec:conclusion} 
In this manuscript we discussed how to provide a comprehensive dynamical description for the quantum measurement process. For the case of projective measures an exhaustive well-established answer is provided by the von Neumann-Ozawa model hinging upon the decoherence induced by an ultimately macroscopic apparatus on the system under investigation. As far as the decoherence process takes place the states of the apparatus, on which the information about measurement outputs is encoded, progressively become orthogonal to each other. Once the decoherence process has taken place such states result to be perfectly distinguishable, thus allowing for an optimal encoding of the measurement results. We proved that this model cannot be directly applied to tackle the case of non-orthogonal measurements, as it could induce a violation of the complete positivity requirement for such dynamical process before the decoherence is completed. We showed different strategies in order to overcome this hindrance. On the one hand it results that it is possible to retrieve the correct probability distribution prescribed by an arbitrary POVM by extending the von Neumann description to an ancillary system and performing a joint projective measure on the system and the ancilla (Appendix~\ref{sec:naim2}). However this solution does not return the expression for the post-measurement  state of the system prescribed by the definition of POVMs. 
In Section~\ref{sec.constr} we show that a possible solution to this problem can be realized by getting rid of such a net separation between the ancilla and the apparatus, and finally identifying the latter with a macroscopic ancilla. The key mechanism underlying our model consists in engineering a coupling between the system and the ancilla in terms of state transfer Hamiltonians acting on orthogonal eigenspaces of the global Hilbert space. By construction this allows to encode the information about the output results of an arbitrary POVM into the states of the ancilla which, at difference with the standard decoherence model, constitute an orthonormal set at all times. This allows us to retrieve not only the correct probability distribution for the output results, but also the correct expression for the post-measurement state of POVMs.

\appendix
\section{CPT conditions for the mapping~(\ref{e.maprho_{_{S}}(t)1})} \label{appax} 

If we force the mapping 
(\ref{e.maprho_{_{S}}(t)1}) to apply also when  the projectors $\ops{\Pi}{\gamma}_{_{S}}$s are  replaced by the element
 $\ops{M}{\gamma}_{_{S}}$ associated to a generic POVM, at local level on $S$ this would induce the following transformation 
\begin{equation}\label{e.maprho_{_{S}}(t)POVM}
\rhoin_{_{S}} \longrightarrow  \sum_{\gamma\gamma'}\ops{M}{\gamma}_{_{S}}\rhoin_{_{S}}{\ops{M} {\gamma'}_{_{S}}}^\dagger
\,_{_{_{\Xi}}}\!\exval{\Xi^{\gamma'}(t)|\Xi^\gamma(t)}_{_{_{\Xi}}}\,.
\end{equation}
Notice that the scalar product $_{_{_{\Xi}}}\exval{\Xi^{\gamma'}(t)|\Xi^\gamma(t)}_{_{_{\Xi}}}$ can be seen as the element $\gamma, \gamma'$ of a positive semidefinite matrix $Q(t)$ in a given orthonormal basis $\{ \ket{\phi_{\gamma}}\}_{\gamma=1, \ldots ,n_\Gamma}$ of a $n_\Gamma$-dimensional Hilbert space.
Let then $Q(t)=\sum_{j=1}^{n_\Gamma} q_j(t) |u_j(t) \rangle \langle u_j(t)|$ be the spectral decomposition of $Q(t)$, with $q_j(t)\geq0$ and 
\begin{eqnarray}
 \sum_{j=1}^{n_\Gamma} q_j(t) = \Tr[Q]= \sum_{\gamma=1}^{n_\Gamma}   \,_{_{_{\Xi}}}\!\exval{\Xi^{\gamma}(t)|\Xi^\gamma(t)}_{_{_{\Xi}}} = 
n_\Gamma\;. \end{eqnarray} 
Writing  $\, _{_{_{\Xi}}}\!\exval{\Xi^{\gamma'}(t)|\Xi^\gamma(t)}_{_{_{\Xi}}}$ in the eigenbasis of $Q(t)$, 
we can then recast  the mapping (\ref{e.maprho_{_{S}}(t)POVM}) 
as
\begin{equation}\label{e.maprho_{_{S}}(t)POVM1}
\rhoin_{_{S}} \longrightarrow
\sum_{j=1}^{n_\Gamma} {q_j(t)} L_{_{S}}^{(j)}(t) \rhoin_{_{S}} {L_{_{S}}^{(j)} (t)}^\dagger,
\end{equation}
where $L_{_{S}}^{(j)}(t) =  \sum_{\gamma=1}^{n_\Gamma}  \bra{u_j(t)}\phi_\gamma \rangle M_{_{S}}^{(\gamma)}$ are operators
 fulfilling the constraint
\begin{eqnarray} \label{CPTCOND1} 
\sum_{j=1}^{n_\Gamma}   {L_{_{S}}^{(j)}(t)}^\dagger L_{_{S}}^{(j)}(t)= \sum_{\gamma}^{n_\Gamma}  
 {\ops{M} {\gamma}}^\dagger_{_{S}} {\ops{M} {\gamma}}_{_{S}} 
 =\openone_{_{S}}\;. 
\end{eqnarray} 
It is then easy to verify that Eq.~(\ref{e.maprho_{_{S}}(t)POVM1}) is CPT if and only if 
the following condition holds
\begin{eqnarray} \label{CPTCOND} 
&&\sum_{j=1}^{n_\Gamma}   {q_j(t)} {L_{_{S}}^{(j)}(t)}^\dagger L_{_{S}}^{(j)}(t)\\
&&\qquad = \sum_{\gamma,\gamma'=1}^{n_\Gamma}  
\,_{_{_{\Xi}}}\!\exval{\Xi^{\gamma'}(t)|\Xi^\gamma(t)}_{_{_{\Xi}}} {\ops{M} {\gamma'}_{_{S}}}^\dagger {\ops{M} {\gamma}_{_{S}}}\nonumber 
 =\openone_{_{S}}\;. 
\end{eqnarray} 
The identity is trivially attained when  the $M_{_{S}}^{(\gamma)}$ form a complete set of orthogonal projectors as in the case of PVMs.
On the contrary if this condition is not met then Eq.~(\ref{CPTCOND}) is in general in conflict with (\ref{CPTCOND1})
with the  exception of the case when the $q_j(t)$ are all equal to 1, forcing $Q(t)$ to be the identity operator, and the vectors
$|\Xi^\gamma(t)\rangle_{_{_{\Xi}}}$ to be orthonormal, i.e. 
 $_{_{_{\Xi}}}\exval{\Xi^{\gamma'}(t)|\Xi^\gamma(t)}_{_{_{\Xi}}}=\delta_{\gamma,\gamma'}$.

\section{$S+A+\Xi$ approach to dynamical mapping}   \label{sec:naim2}

A reasonable, yet non completely satisfying approach, to produce a generic dynamical model for describing an arbitrary POVM follows by considering
the extended $SA$ system of the Naimark representation as the system of interest, and introducing 
an external environment $\Xi$ which perform a PVM on it. 
 First we notice that any PVM $\{\ops{\Pi}{\gamma}_{_{S\!A}}\}$ on $SA$ together with an arbitrary state
$\proj{\psi_0}{A}$, defines
a POVM on $S$ with measurement operators 
$\{\ops{F}{\gamma}_{_{S}}{=_{_{_A}}\!}\exval{\psi_0|\ops{\Pi}{\gamma}_{_{S\!A}}|\psi_0}\!_{_{_A}}\}$. 
Actually, thanks to the Naimark theorem, the reverse statement is also true.
Indeed, if we take an arbitrary POVM $\{\ops{F}{\gamma}_{_{S}}\}$ on $S$, from Eqs.~\eqref{e.E_gamma-Naimark} and \eqref{e.p_gamma-Naimark} 
we can define the projectors
\begin{equation}
\ops{P}{\gamma}_{_{S\!A}}:=V_{_{S\!A}}^\dagger{\mathbb 
I}_{_{S}}\otimes\proj{\gamma}{A}V_{_{S\!A}}~,
\label{e.setPigamma}
\end{equation}
which form a complete orthonormal set in the space  ${\cal L}({\cal H}_{_{S\!A}})$ of linear operators of $S+A$. 
Let us now construct the vN-O dynamical model for such
PVM introducing the  interaction $O_{_{S\!A}}\otimes O_{_{\Xi}}$, 
with $O_{_{S\!A}}=\sum_\gamma o_\gamma \ops{P}{\gamma}_{_{S\!A}}$, 
$o_\gamma\in{\mathbb R}$, and $O_{_{\Xi}}$ self-adjoint;
the corresponding propagator reads
\begin{equation}\label{eq:dynam_POVM}
U_{_{S\!A\Xi}}(t) := e^{-i tO_{_{S\!A}}\otimes O_{_{\Xi}} }= \sum_\gamma 
\ops{P}{\gamma}_{_{S\!A}}\otimes \ops{U}{\gamma}_{_{\Xi}}(t)~,
\end{equation}
with $\ops{U}{\gamma}_{_{\Xi}}(t)=e^{-i t o_\gamma  O_{_{\Xi}} }$.  Subject to such unitary, an initial state 
$\rhoin_{_{S}}\otimes\proj{\psi_0}{A}\otimes\proj{D}{\Xi}$
evolves into  
\begin{equation}
\rho_{_{SA\Xi}}(t) =
\sum_{\gamma, \gamma'}\ops{P}{\gamma}_{_{S\!A}}\rhoin_{_{S}}\otimes\proj{\psi_0}{A}
\ops{P}{\gamma'}_{_{S\!A}} \otimes \ket{\Xi^\gamma(t)}_{\!_{_{\Xi}}\!} \bra{\Xi^{\gamma'}(t)}\,,
\nonumber
\end{equation}
at a later time $t$, and for $t > t_{\rm d}$  the density operator of the joint system $SA$ 
will have a block-diagonal form with respect to the basis of the PVM $\{\ops{P}{\gamma}_{_{S\!A}}\}$.
From the viewpoint of the principal system $S$, the composite system $A+\Xi$ is however seen as a single measurement apparatus: In this perspective, if we expand $\rho_{_{SA\Xi}}(t)$ into an arbitrary basis $\{ \ket{e_k}_{\!_{_A}}\}$ of ${\cal H}_{_A}$, we have
\begin{eqnarray}\label{e.xia}
\rho_{_{SA\Xi}}(t)&=&\sum_{\gamma, \gamma'} {_{\!_{_A}\!}}\bra{e_k}\ops{P}{\gamma}_{_{S\!A}}|\psi_0\rangle_{_A} \;  \rhoin_{_{S}} \;
{_{_A} \langle} \psi_0| 
\ops{P}{\gamma'}_{_{S\!A}} \ket{e_{k'}}_{\!_{_A}} \nonumber\\ &&\qquad \ket{e_k}_{\!_{_A}\!} \bra{e_{k'}}\otimes \ket{\Xi^\gamma(t)}_{\!_{_{\Xi}}\!} \bra{\Xi^{\gamma'}(t)}\,,
\end{eqnarray}
and 
\begin{equation}
\rho_{_{S\!A}}(t)=
\sum_\gamma\ops{P}{\gamma}_{_{S\!A}}\rhoin_{_S}\otimes\proj{\psi_0}{A}
\ops{P}{\gamma}_{_{S\!A}}=\sum_\gamma 
p_\gamma\ops{\rho}{\gamma}_{_{S\!A}}=\rhoout_{_{S\!A}}\,.
\label{e.rho_SA(tau)}
\end{equation}
Therefore, the system experiences a decoherence process which takes place in $n_\Gamma$  ($n_{_A}$-dimensional) subspaces spanned by $\{\ket{e_k}_{_A} \ket{\Xi^\gamma(t)}_{_{\Xi}}\}_{k=1, \ldots n_{_A}}$ of ${\cal H}_{_{A \Xi}}$. Indeed, since just the vectors $\{\ket{\Xi^\gamma(t)}_{_{\Xi}}\}$ evolve in time, being $\Xi$ the actual macroscopic part of the apparatus, the effective decoherence process will emerge only with respect to the label $\gamma$. 
As we are aiming at a dynamical model for the original POVM 
on $S$, the relevant question is: what happens at the level of the principal system $S$? Clearly, no matter which subsystem we are going to identify as
the apparatus (say $\Xi$ or $\Xi +A$)  Eq.~(\ref{e.xia}) has not the form we are aiming to, not
yielding to something like (\ref{e.maprho_{_{S}}(t)1POVM}) even after the orthogonalization of the $\ket{\Xi^{\gamma}(t)}_{_\Xi}$s. 
As for the probability distribution $\{p_\gamma\}$, the outcomes statistics generated by 
$\{F_{_S}^{(\gamma)}\}$ via 
$p_\gamma=\Tr[\rhoin_{_{S}}\ops{F}{\gamma}_{_{S}}]$ is nevertheless the same as that entering 
Eq.~\eqref{e.rho_SA(tau)}, as can be easily seen by 
Eqs.~\eqref{e.E_gamma-Naimark} and \eqref{e.setPigamma}.
As for the state of $S$, by inserting the explicit expression for $P_{_{SA}}^{(\gamma)}$ into \eqref{e.setPigamma} and tracing over the ancilla, we get 
\begin{eqnarray}\label{e.rhoStauW}
\rho_{_{S}}(t)&=&\sum_{\gamma k}
N_{_{S}}^{(k){\gamma}}
\left(M_{_{S}}^{(\gamma)} \rhoin_{_{S}}  {M_{_{S}}^{(\gamma)}}^\dagger\right)
{N_{_{S}}^{(k){\gamma}}}^\dagger\nonumber\\
&=&\sum_\gamma p_\gamma 
\sum_k N_{_{S}}^{(k){\gamma}}
\ops{\rho_{_{S}}}{\gamma} {N_{_{S}}^{(k){\gamma}}}^\dagger
\end{eqnarray}
where $N_{_{S}}^{(k){\gamma}}:= {_{_A}}\!\exval{e_k|V^\dagger_{_{S\!A}}|\gamma}\!_{_{_A}}$. Therefore, $\rho_{_{S}}(t)$ does not coincide with the post-measurement state $\rhoout_{_{S}}$ defined in Eq.~\eqref{e.rhoout}.  The only exception is represented by the case in which $\dim{\cal H}_{_{S}}=n_{\Gamma}=n_{_A}$ and $V_{_{SA}}$ coincides with the {\it swap} operator $\mathbb{S}_{S A}:= 
\sum_{\gamma,\gamma'} |\gamma\rangle_{\!_{_{S}} \!}\langle \gamma'| \otimes 
|\gamma'\rangle_{\!_{_A} \!}\langle \gamma|$: in this case it results $P_{_{SA}}^{(\gamma)}=|\gamma\rangle_{_{S}}\langle \gamma| \otimes \mathbb{I}_{_A}$ , which pulls back to the vN-O model for the ideal PVM $\{ |\gamma\rangle_{\!_{_{S}}\!}\langle \gamma| \}$ on $S$.
(The operator $\protect \mathbb  {S}_{S A}$ is a unitary self-adjoint transformation such that for all operators $\Theta _{_{S}} \in {\protect \cal  L}({\protect \cal  H}_{_{S}})$ and $\Upsilon _{_A} \in {\protect \cal  L}({\protect \cal  H}_{_A})$ gives $\protect \mathbb  {S}_{_{SA}}(\Theta _{_{S}}\otimes \Upsilon _{_A})\protect \mathbb  {S}_{_{SA}}=\Theta _{_A}\otimes \Upsilon _{_{S}}$). 

\begin{figure}[t!]
\includegraphics[width=0.21\textwidth]{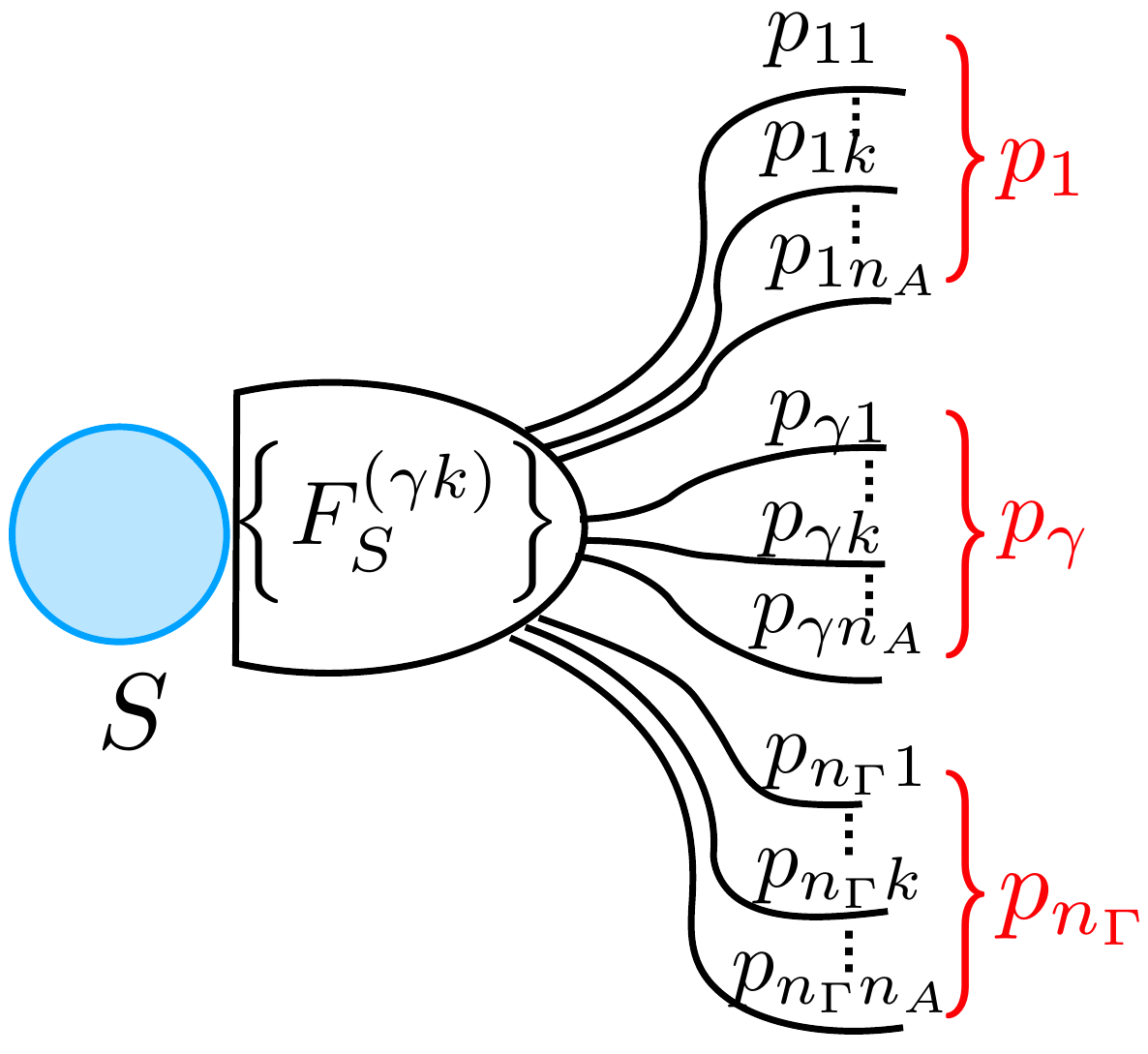}
\caption{Correspondence between the statistics $\{p_\gamma\}$ yielded by 
the POVM $\{F_{_{S}}^{(\gamma)}\}
{=}\{\hspace*{-0.4mm}_{_A} \langle 
\psi_0 | P_{_{S\!A}}^{(\gamma)} | \psi_0\rangle_{_A}\}$ and the one resulting 
from $\{ F_{_{S}}^{(\gamma k)} \}$.}
\label{fig:generalized_meas}
\end{figure}

However, we can push forward. Let us observe that for any fixed $\gamma$ the set of operators $\{N_{_{S}}^{(k){\gamma}\dagger}N_{_{S}}^{(k){\gamma}}\}$ returns a resolution of the identity, i.e $\sum_{k} N_{_{S}}^{(k){\gamma}\dagger}N_{_{S}}^{(k){\gamma}} = \mathbb{I}_{_{S}}$.  From this, two facts follow: The first is that $\rho_{_{S}}(t)$ reads as the post-measurement state of a double-labeled POVM $\{F_{_{S}}^{(\gamma,k)}\}$, with measurement operators $\{M_{_{S}}^{(\gamma k)}:=N_{_{S}}^{(k){\gamma}} M_{_{S}}^{(\gamma)}\}$. Such POVM accounts for $n_\Gamma n_{_A}$  possible outcomes, and the associated probability distribution  
$p_{\gamma k}=\Tr[\rhoin_{_{S}} \ops{F}{\gamma k}_{_{S}}]$ is related to that of the original POVM $\{\ops{F}{\gamma}_{_{S}}\}$ via
\begin{equation}
\sum_k p_{\gamma k}=p_\gamma~.
\end{equation}
This means that, if we gather the $n_\Gamma {\cdot} n_{_A}$  
outcomes $\mu_{\gamma k}$ of the POVM $\{\ops{F}{\gamma k}_{_{S}}\}$
in $n_\Gamma$ sets ${\cal O}_\gamma$, each 
bearing $n_{_A}$ elements, 
${\cal O}_\gamma=\{\mu_{\gamma 1},\mu_{\gamma 2},...
\mu_{\gamma n_{_A}}\}$ 
(see Figure~\ref{fig:generalized_meas}), the probability for each set is
the sum of the probabilities for the outcomes it collects. This is consistent with the fact that, as observed through Eq.\eqref{e.xia}, from the viewpoint of the principal system, the decoherence process emerges in the form of $n_\Gamma$ subspaces (one for each $\gamma$) in ${\cal H}_{A \Xi}$. 

The second fact following from the condition $\sum_{k} N_{_{S}}^{(k){\gamma}\dagger}N_{_{S}}^{(k){\gamma}} = \mathbb{I}_{_{S}}$ is that for all $\gamma$s, the set $\{N_{_{S}}^{(k){\gamma}\dagger}N_{_{S}}^{(k){\gamma}}\}$ itself defines a POVM on ${\cal H}_{_{S}}$. This represents a meaningful result, as it tells us that the state ~\eqref{e.rhoStauW} prior 
to the output production is the statistical mixture, with the 
original POVM's probability distribution $\{p_\gamma\}$, of the $n_\Gamma$ output
states of a set of non-selective measurements, each labeled by 
$\gamma$ and defined by the set of measurement operators $\{N_{_{S}}^{(k){\gamma}}\}$,  
performed upon the respective $\gamma$-detected state $\rho_{_{S}}^{(\gamma)}$ resulting 
from the action of the original POVM on $\rhoin_{_{S}}$.

\end{document}